\documentclass[conference,final]{IEEEtran}

\usepackage{booktabs} %
\usepackage{listings}

\usepackage{graphicx}
\usepackage{amsmath}
\usepackage{amssymb}
\usepackage{color}
\usepackage{ifpdf}
\usepackage{float}
\usepackage[utf8]{inputenc}
\usepackage{multirow}
\usepackage{rotating}
\usepackage{subfigure}
\usepackage{setspace}
\usepackage{amsmath}
\usepackage{moresize}
\usepackage{url}
\usepackage{booktabs}
\usepackage{listings}
\usepackage{paralist}
\usepackage{wrapfig}
\usepackage{multirow}
\usepackage{ifpdf}
\usepackage{xspace}
\usepackage{keyval}
\usepackage{color}

\definecolor{listinggray}{gray}{0.95}
\definecolor{darkgray}{gray}{0.7}
\definecolor{commentgreen}{rgb}{0, 0.4, 0}
\definecolor{darkblue}{rgb}{0, 0, 0.4}
\definecolor{middleblue}{rgb}{0, 0, 0.7}
\definecolor{darkred}{rgb}{0.4, 0, 0}
\definecolor{brown}{rgb}{0.5, 0.5, 0}

\usepackage[normalem]{ulem}
\makeatletter
\def\cyanuwave{\bgroup \markoverwith{\lower3.5\p@\hbox{\sixly \textcolor{cyan}{\char58}}}\ULon}
\def\reduwave{\bgroup \markoverwith{\lower3.5\p@\hbox{\sixly \textcolor{red}{\char58}}}\ULon}
\def\blueuwave{\bgroup \markoverwith{\lower3.5\p@\hbox{\sixly \textcolor{blue}{\char58}}}\ULon}
\font\sixly=lasy6 %
\makeatother

\usepackage{xcolor}
\usepackage[colorlinks]{hyperref}
\AtBeginDocument{%
  \hypersetup{
    citecolor=blue,
    linkcolor=blue,   
    urlcolor=blue}}

\newif\ifdraft
\ifdraft
\definecolor{ocolor}{rgb}{1,0,0.4}
\newcommand{\onote}[1]{ {\textcolor{ocolor} { (***Ole: #1) }}}
\newcommand{\terminology}[1]{ {\textcolor{red} {(Terminology used: \textbf{#1}) }}}

\newcommand{\jhanote}[1]{ {\textcolor{red} { ***shantenu: #1 }}}
\newcommand{\alnote}[1]{ {\textcolor{blue} { ***andreL: #1 }}}
\newcommand{\amnote}[1]{ {\textcolor{blue} { ***andreM: #1 }}}
\newcommand{\georgenote}[1]{ {\textcolor{brown} { ***sharath: #1 }}}
\newcommand{\revThreeNote}[1]{ {\textcolor{purple}{}}} 
\newcommand{\revOneNote}[1]{ {\textcolor{purple}{}}} 
\newcommand{\revTwoNote}[1]{ {\textcolor{purple}{}}} 
\definecolor{orange}{rgb}{1,.5,0}
\newcommand{\aznote}[1]{ {\textcolor{orange} { ***ashley: #1 }}}
\definecolor{dandelion}{cmyk}{0,0.29,0.84,0}
\newcommand{\mtnote}[1]{ {\textcolor{dandelion} { ***matteo: #1 }}}
\newcommand{\note}[1]{ {\textcolor{magenta} { ***Note: #1 }}}
\else
\newcommand{\onote}[1]{}
\newcommand{\terminology}[1]{}

\newcommand{\alnote}[1]{}
\newcommand{\amnote}[1]{}
\newcommand{\athotanote}[1]{}
\newcommand{\georgenote}[1]{}
\newcommand{\pmnote}[1]{}
\newcommand{\jhanote}[1]{}
\newcommand{\msnote}[1]{}
\newcommand{\mrnote}[1]{}
\newcommand{\aznote}[1]{}
\newcommand{\mtnote}[1]{}
\newcommand{\note}[1]{}
\newcommand{\revOneNote}[1]{}
\newcommand{\revTwoNote}[1]{}
\newcommand{\revThreeNote}[1]{} 
\fi

\newcommand{\pilot}{Pilot\xspace}

\newcommand{\pilotjob}{Pilot-Job\xspace}
\newcommand{\pilotjobs}{Pilot-Jobs\xspace}

\newcommand{\computeunit}{Compute-Unit\xspace}
\newcommand{\computeunits}{Compute-Units\xspace}

\newcommand{\upp}{\vspace*{-0.5em}}

\lstdefinestyle{myListing}{
  frame=single,
  backgroundcolor=\color{listinggray},
  language=C,
  basicstyle=\ttfamily \footnotesize,
  breakautoindent=true,
  breaklines=true
  tabsize=2,
  captionpos=b,
  aboveskip=0em,
  belowskip=-2em,
}

\lstdefinestyle{myPythonListing}{
  frame=single,
  backgroundcolor=\color{listinggray},
  language=Python,
  basicstyle=\ttfamily \scriptsize,
  breakautoindent=true,
  breaklines=true
  tabsize=2,
  captionpos=b,
}

\ifpdf
\DeclareGraphicsExtensions{.pdf, .jpg, .tif}
\else
\DeclareGraphicsExtensions{.ps,  .eps, .jpg}
\fi

\usepackage{listings}
\usepackage{paralist}
\lstnewenvironment{code}[1][]%
{
\noindent
\minipage{1.0 \linewidth}
\vspace{0.5\baselineskip}
\lstset{
    language=Python,
    frame=single,
    captionpos=b,
    stringstyle=\ttfamily,
    basicstyle=\scriptsize\ttfamily,
    showstringspaces=false,#1}
}
{\endminipage}

\defaultleftmargin{1em}{}{}{}
\begin{document}

\title{Pilot-Streaming: A Stream Processing Framework
  for High-Performance Computing\upp}

\author{Andre Luckow$^{1,2,3}$, George Chantzialexiou$^{1}$, Shantenu Jha$^{1,4}$\\
   {\footnotesize{\emph{$^{1}$RADICAL, ECE, Rutgers University, Piscataway,NJ 08854, USA}}}\\
   \footnotesize{\emph{$^{2}$Clemson University, Clemson, SC 29634, USA}}\\
   \footnotesize{\emph{$^{3}$Ludwig Maximilian University, Munich, Germany}}\\
   \footnotesize{\emph{$^{4}$Brookhaven National Laboratory, Upton, NY, USA}\upp\upp\upp}
   }

\date{}
\maketitle

\begin{abstract}

An increasing number of scientific applications utilize stream processing to
analyze data feeds of scientific instruments, sensors, and simulations. In
this paper, we study the streaming and data processing requirements of light
source experiments, which are projected to generate data at 20 GB/sec in the
near future. As beamtimes available to users are typically short, it is
essential that processing and analysis can be conducted in a streaming mode.
The development and deployment of streaming applications is a complex task and
requires the integration of heterogeneous, distributed infrastructure,
frameworks, middleware and application components written in different
languages and abstractions. Streaming applications may be extremely dynamic
due to factors, such as variable data rates, network congestions, and
application-specific characteristics, such as adaptive sampling techniques and
the different processing techniques. Consequently, streaming system are often
subject to back-pressures and instabilities requiring additional
infrastructure to mitigate these issues. We propose \emph{Pilot-Streaming}, a
framework for supporting streaming applications and their resource management
needs on HPC infrastructure. Underlying Pilot-Streaming is a unifying
architecture that decouples important concerns and functions, such as message
brokering, transport and communication, and processing. Pilot-Streaming
simplifies the deployment of stream processing frameworks, such as Kafka and
Spark Streaming, while providing a high-level abstraction for managing
streaming infrastructure, e.\,g.  adding/removing resources as required by the
application at runtime. This capability is critical for balancing complex
streaming pipelines. To address the complexity in the development of streaming
applications, we present the Streaming Mini-Apps, which supports different
plug-able algorithms for data generation and processing, e.\,g., for
reconstructing light source images using different techniques. We use the
streaming Mini-Apps to evaluate the Pilot-Streaming framework demonstrating
its suitability for different use cases and workloads.

\end{abstract}

\section{Introduction}

Stream processing capabilities are increasingly important to analyze and
derive real-time insights on incoming data from experiments, simulations, and
Internet-of-Things (IoT) sensors~\cite{streaming2015}. Prominent examples are
synchrotron light source experiments, such as those at the National
Synchrotron Light Sources II (NSLS-II) or the X-Ray Free Electron Laser (XFEL)
light sources. Some experiments at these light sources are projected to
generate data at rates of 20\,GB/sec~\cite{nsls}. This data needs to be
processed in a time-sensitive if not real-time manner, to support steering of
the experiments~\cite{lcls_data}.

Further, an increasing number of scientific workflows integrate simulations
either with data from experimental and observational instruments, or conduct real-time analytics of simulation data~\cite{Malakar:2016:OEC:3014904.3014985}.
Workflows are stymied by the fact that capabilities to  continuously process
time-sensitive data on HPC infrastructures are underdeveloped while they
require sophisticated approaches for resource management, data movement and
analysis. The complex application and resource utilization patterns of
streaming applications critically demand dynamic resource management
capabilities. For example, minor changes in data rates, network bandwidths, and
processing algorithms can lead to imbalanced and dysfunctional system.

\jhanote{I removed abstractions as I think the claim is that Pilot-Streaming
is a framework and not the abstraction. The definition needs fixing as we now
we have a framework for a framework ... }\alnote{makes sense}

We propose \emph{Pilot-Streaming}, a framework designed to efficiently deploy
and manage streaming frameworks for message brokering and processing, such as
Kafka~\cite{kreps2011kafka}, Spark~\cite{Zaharia:2010:SCC:1863103.1863113} and
Dask~\cite{dask}, on HPC systems. Underlying Pilot-Streaming is a unifying
architecture that decouples important concerns and functions, such as message
brokering, transport and communication, and processing. Pilot-Streaming is
based on the Pilot-Job concept and the Pilot-Abstraction~\cite{pstar12}.
Pilot-Streaming enables application and middleware developers
\jhanote{developers of what -- applications? middleware? frameworks?}
\alnote{refined. addressed both applications and middleware devs} to deploy,
configure and manage frameworks and resources for complex streaming
applications. Acquired resources can be dynamically adjusted at runtime -- a
critical capability for highly dynamic streaming applications. Further,
Pilot-Streaming serves as unifying \jhanote{Unify is typically a binary (or
n-ary) operator, so unify X with Y. What might be X and Y. Do you mean "common"
or "integrative"?} \alnote{} 
API layer for managing computational tasks in an interoperable, 
framework-agnostic way, i.\,e. it allows the implementation of streaming 
tasks that can run both in Spark Streaming, Dask or other frameworks.

\jhanote{This is back to the "problem space" -- which is what the 2nd
paragraph was about. Maybe flesh out / finalize problem space before moving to
solution space, i.e., pilot-streaming?} \alnote{shortened intro to Mini-Apps to remove redundancy. }

To further address the development and deployment challenges of streaming apps,
we develop the \emph{Streaming Mini-Apps} framework based on a systematic
analysis of different scientific streaming application~\cite{sminiapps}. The
Mini-Apps provides the ability to quickly develop streaming applications and to
gain an understanding of the performance of the pipeline, existing bottlenecks,
and resource needs. We demonstrate the capabilities of Pilot-Streaming and the
Streaming Mini-Apps by conducting a comprehensive set of experiments evaluating
the processing throughput of different image reconstruction algorithms used in
light source sciences.

This paper makes the following contributions: (i) It surveys the current state
of message broker and streaming frameworks and their ability to support
scientific streaming applications; (ii) \jhanote{shouldn't this reference the
mini-apps as the realization of the conceptual framework?}\alnote{done} It
provides a conceptual framework for analyzing scientific streaming applications
and applies it to a machine learning and light source analytics use case.
The Mini-App framework provides a simple solution for simulating
characteristics of these applications. (iii) It presents an abstraction and
architecture \jhanote{both?}\alnote{actually yes sec II has the normative
architecture and IV the specific instantiation via PS} for stream processing on
HPC. Pilot-Streaming is a reference implementation of that architecture, and
(iv) It demonstrates and evaluates the described capabilities 
using a set of large-scale experiments on the XSEDE machine Wrangler, for
streaming machine learning and different light source reconstruction
algorithms. \jhanote{Isn't there overlap between points (ii) and (iv) when
discussing the use case?}\alnote{distinguished ii and iv better. Another option would be to move iii to ii and combine ii and iv (use case + evaluation)}

This paper is structured as follows: In Section~\ref{sec:background_related}
we investigate the architectural components of a typical streaming
infrastructures and applications and related work. We continue with an
analysis of streaming applications in Section~\ref{sec:applications}.
Section~\ref{sec:arch_impl} presents the architecture, capabilities and
abstractions provided by Pilot-Streaming. The frameworks serves as basis for
the Mini-Apps discussed in Section~\ref{sec:miniapps}. In
Section~\ref{sec:exp} we present an experimental evaluation of
Pilot-Streaming.

\section{Background and Related Work}
\label{sec:background_related}

\emph{We define a streaming application as an application that processes and
acts on an unbounded stream of data close to real time.} In this section we
describe the current state of streaming middleware and infrastructure and
related work. There is no
consensus on software and hardware infrastructure for streaming applications,
which increases the barrier for adoption of streaming technology in a broader
set of application (see Fox et al.~\cite{streaming2015}). Notwithstanding the lack of  consensus, in this
paper we will explore the usage of the existing \pilot-Abstractions as a
unified layer for the development of streaming applications.

\subsection{Streaming Middleware and Infrastructure}

\alnote{Reviewer: Table I is not mentioned in the corresponding section (II A Streaming Processing Frameworks).}\alnote{done}

\begin{figure}[t]
  \centering
    \includegraphics[width=.49\textwidth]{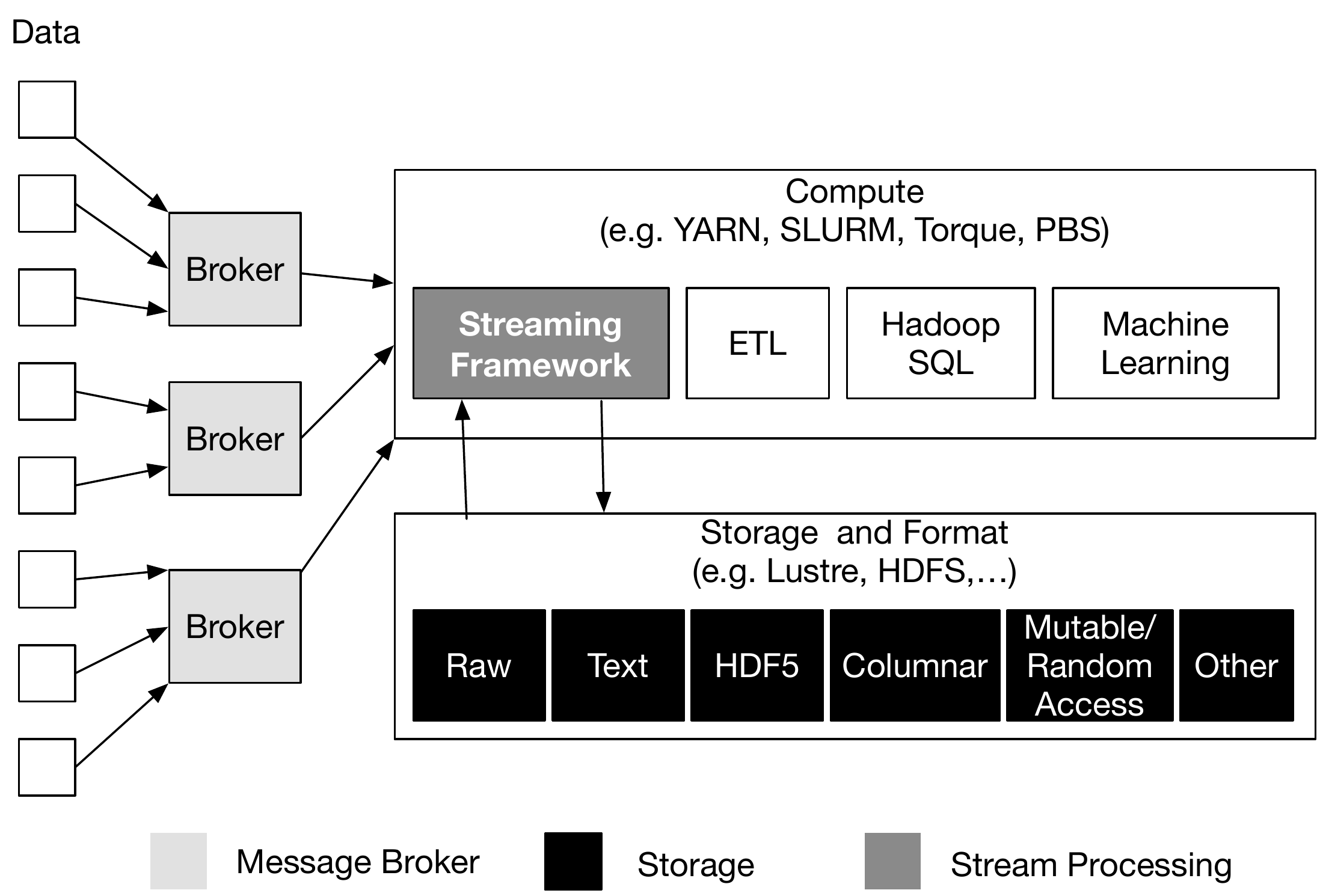}
  \caption{\textbf{Streaming Applications Architecture:} The \emph{message broker} decouples streaming applications from incoming data feeds and enables multiple applications to process the data. The \emph{streaming framework} typically provides a windowing abstraction on which
  user-defined functions can be performed.}
  \label{fig:figures_streaming}
\end{figure}

The landscape of tools and frameworks for stream processing is heterogeneous
(see~\cite{fox-streaming-2016} for survey). Figure~\ref{fig:figures_streaming}
illustrates the main components of a stream system are: the message broker, the
storage and the stream processing engine. We will investigate these in the
following section.

\textbf{Message Broker:} The broker decouples data producers and consumers
providing a reliable data storage and transport. By combining data transport
and storage, the message broker can provide a durable, replay-able data source
to streaming processing applications. For this purpose, the brokering system
typically provides a publish-subscribe interface. The best throughputs are
achieved by log-based brokering systems, such as
Kafka~\cite{journals/pvldb/WangKSPZNRKS15}. Facebook Logdevice~\cite{logdevice}
provides a similar log abstraction, but with a richer API (record not byte
based) and improved availability guarantees. Apache Pulsar is another
distributed brokering system~\cite{pulsar}. Other types of publish-subscribe
messaging system exist, such as ActiveMQ and RabbitMQ, but are generally less
scalable than distributed log-based services, such as
Kafka~\cite{kreps2011kafka}. A message broker enables application to observe a
consistent event stream of data at its own pace executing complex analytics on
that data stream. Kafka is one such distributed message broker optimized for
large volume log files containing event streams of data. Amazon
Kinesis~\cite{kinesis} and Google Cloud Pub-Sub~\cite{google_pubsub} are two
 message brokers offered as ``platform as a service'' in the cloud.

\begin{table}[t]
	\scriptsize
	\begin{tabular}{|p{1.2cm}|p{1.5cm}|p{1.6cm}|p{1.4cm}|p{1.4cm}|}\hline
	& \textbf{Storm/Heron} & \textbf{Spark Streaming} & \textbf{Flink} & \textbf{Dask Streamz} \\\hline
	Description & Java/C++ with Python API & Scala with Java, Python APIs & Java & Python \\\hline
	Architecture & Continuous & Mini-batch, Continuous & Continuous & Mini-batch \\\hline
	Windowing & Yes & Event time introduced with structured API & Yes with event/processing time & Fixed Time \\\hline
	Higher-Level APIs & Streamlet API
	(MapReduce) & Structured Streaming (DataFrames, SQL) & Data Tables & DataFrames (state-less) \\\hline
	Guarantees & Exactly once & Exactly once & Exactly once & No \\\hline
	Integration & Kafka & Kafka, Kinesis & Kafka & Kafka \\\hline
	\end{tabular}
	\caption{\textbf{Stream Processing Frameworks~\label{tab:stream_processing}}\upp\upp\upp\upp\upp}
\end{table}

\textbf{Streaming Processing Frameworks:} 
A heterogeneous landscape of infrastructures and
tools supporting streaming needs on different levels emerged.
Table~\ref{tab:stream_processing} summarizes the properties of four important
stream processing systems. Batch frameworks, such as
Spark~\cite{Zaharia:2010:SCC:1863103.1863113} and Dask~\cite{dask}, have been
extended to provide streaming
capabilities~\cite{Zaharia:2013:DSF:2517349.2522737,streamz}, while different
native streaming frameworks, such as Storm~\cite{storm}, Heron~\cite{heron} and
Flink~\cite{flink} have emerged. Apache Beam~\cite{beam} is high-level
streaming abstraction that can be used together with Flink and Spark and is
available as managed cloud service called Google
Dataflow~\cite{google-dataflow}. Apache Beam's abstraction is based on a
rigorous model and provides well-defined and rich semantics for windowing,
transformations and other operations. The different stream processing engines
differs significantly in the ways they handle events and provide processing
guarantees: Storm and Flink continuously process data as it arrives. Dask
Streamz and Spark Streaming rely on micro-batches, i.\,e.,\ incoming data is
partitioned into batches according to a user-defined criteria (e.\,g.\ time
window). The advantage of micro-batching is that it provides better fault
tolerance, higher throughput and exactly-once processing guarantees, while 
native stream engines can provide lower latencies and more advanced windowing 
capabilities, e.\,g., tumbling and session-based windows.

Each of the described message brokers and stream processing frameworks provides
unique capabilities, e.\,g., specific windows semantics, high-level APIs (such
as streaming SQL), low latency. However, they do not address interoperability,
deployment on HPC and resource management. While all frameworks provide an
application-level scheduler, resource management is typically a second-order
concern and not addressed in a generalized, holistic, framework-agnostic
approach.

\revThreeNote{In Section 2.1 authors argue that “brokers and streaming engines do not address interoperability, deployment on HPC and resource management” and that resource management is typically a second-order concern and not addressed in a generalized, holistic, framework agnostic approach. There is previous work on integrating Spark on HPC; how does the current work compare with those?
}\alnote{Spark on HPC is addressed now in section II.B}

\subsection{Related Work}

There are several areas of related work: (i) frameworks that allow the
interoperable use of streaming frameworks on HPC, (ii) the usage of HPC
hardware features and frameworks (such as MPI) to optimize data streaming
frameworks, and (iii) the exploration of data streaming in distributed
applications.

{\bf Interoperable Streaming on HPC:}
Various tools have been proposed to support open source Big data frameworks,
such as Hadoop and Spark on HPC environments on top of schedulers like SLURM,
PBS/Torque etc~\cite{Krishnan04myhadoop,spark-on-hpc}. Other more
streaming-oriented frameworks, such as Flink, Heron and Kafka are not supported
on HPC out-of-the-box and require the manual implementation of job submission
scripts.

While these script-based approaches is acceptable for small applications, it has
severe limitations with respect to maintainability and support for more complex
stream processing landscapes. For example, it is typically necessary to
coordinate resources among several tools and frameworks, such as simulation and
data acquisition, data message broker, and the actual stream processing
framework. Also, streaming application are much more dynamic exhibiting varying
data production and process rates, than traditional simulation and data
analytics applications. Thus, in this paper we propose the usage of the
Pilot-Abstraction as unifying layer for managing a diverse set of resources and
stream processing frameworks.

\emph{\bf Optimizing Streaming on HPC:} The ability to leverage HPC hardware and
software capabilities to optimize Big Data frameworks has been extensively
explored. Kamburugamuve et al.~\cite{kamburugamuve2016towards} propose the
usage of optimized HPC algorithms for low-latency communication (e.\,g.\ trees)
and scheduling of tasks to enhance distributed stream processing in the Apache
Storm framework~\cite{storm}. In~\cite{kamburugamuve2017} they investigate the
usage of HPC network technology, such as Infiniband and Omnipath, to optimize
the interprocess communication system of Heron~\cite{heron}, the successor of
Storm. Chaimov et al.~\cite{Chaimov:2016:SSH:2907294.2907310} propose the usage
of a file pooling layer and NVRAM to optimize Spark on top of Lustre
filesystems. These approaches can complimentary to the high-level resource
management approach proposed in this paper and can be used to optimize critical
parts of a stream processing pipeline. These approaches mainly focus on 
low-level optimization of Big Data frameworks for HPC. Pilot-Streaming address 
critical gaps in the integration of these frameworks with the application and 
the ability to manage resources across these frameworks in a high-level and 
uniform way.

\emph{\bf Streaming in Scientific Application:} Fox et
al.~\cite{fox-streaming-2016} identifies a broad set of scientific applications
requiring streaming capabilities. Many aspects of these use cases have been
explored: For example, Bicer et al.~\cite{8109123} investigates different light
source reconstruction techniques on HPC. Du~\cite{7938385} evaluates streaming
infrastructure for connected vehicle applications. Both approaches focus solely
on a specific aspect of a single use cases, e.\,g., latencies or processing
throughput. Proving a generalized architecture and solution for many use cases
addressing important shared concerns, such as resource management, is not in
scope of these approaches. Pilot-Streaming and the Streaming Mini-Apps provide
a holistic approach for addressing a broad set of use cases end-to-end from
data source, broker to processing on heterogeneous infrastructure.

The implementation of scientific streaming applications requires the
integration of infrastructure, a diverse set of frameworks: from resource
management, message brokering, data processing to advanced analytics. In most
cases, the data source is external making it essential for streaming
application to dynamically manage resources and frameworks. 

\section{Streaming Applications}
\label{sec:applications}

Stream processing is becoming an increasingly important for
scientific applications. While many streaming applications primarily
perform simple analytics (smooth averages, max detection) on the incoming
data, the computational demands are growing. For example, to run complex reconstruction algorithms for light source data streams or deep
learning based computer-vision algorithms, such as convolutional neural
networks, a vast amounts of scalable compute
resources are required. In this section, we develop a taxonomy for classifying
streaming applications. Further, we will discuss light
source streaming as specific applications example.

\subsection{Applications Characteristics}

In the following we investigate different types of streaming applications in
particular with respect to types data production (simulation, experiment) and
processing:

\textbf{Type 1 -- Experimental Data Streaming:} Experimental data
  generated by an instrument that is processed by a data analysis application
  and/or a simulation. An example are light source experiments (see
  section~\ref{sec:lightsource}).
	
\textbf{Type 2 -- Simulation Data Streaming:} Simulation produces
	data that is processed by a data analysis application. This form of
	processing is referred to as in-situ processing. Different forms of in-situ
	analysis exist: the analysis tasks can e.\,g.\ run within the same HPC job
	or on a separate set of nodes coupled via shared storage and/or network. An 
	example of co-analysis of molecular dynamics 
	simulations data~\cite{Malakar:2016:OEC:3014904.3014985}.
	
\textbf{Type 3 -- Streaming with Feedback/Control Loop:} Data is processed with realtime feedback, i.\,e. output is used to steer simulation respectively experiment. Both type 1 and 2 applications typically benefit from the ability
to integrate real-time insights into an experiment or simulation run.

Streaming applications involve the coupling a data source (simulation,
experimental instrument), message broker and processing. In
general, these components can be deployed  across heterogeneous, distributed
infrastructure. Often, it makes sense to run some pre-processing close to
the datasource (on the edge), transmit selected data to the cloud and do global
processing in the cloud. Resource needs are highly
dynamic and can change at runtime. Thus, an in-depth understanding of application and infrastructure characteristics is required.

The coupling between data source and processing can be (i) direct (e.\,g.,
using a direct communication channel, such as memory) or (ii) indirect via a
brokering system. The direct couple is used when low latencies and realtime
guarantees are required. The direct coupling approach is associated
with several drawbacks: it involves a large amounts of custom code
for interprocess communication, synchronization, windowing, managing data flows
and different data production/consumptions rates (back-pressure) etc. Thus, it
is in most cases advantageous to de-couple production and consumption using a
message broker, such as Kafka. Another concern is the geographic distribution
of data generation and processing: both can be co-located or geographically 
distributed. Further, the number of producer and consumers can vary.

The third component is the actual stream data processing: in simply cases the
application utilizes non-complex analytics on the incoming data, e.\,g.\ for
averaging, scoring, classification or outlier detection. Typically, streaming
applications utilize less complex analytics and operate on smaller amounts of
data, a so-called streaming window. There are multiple types of windowing, e.\,g.\ a fixed, sliding or session  window. Commonly the streaming windows is
either defined based on processing time or event time. More complex application
involve combine analytics with state and model updates, e.\,g.\ the update of a
machine learning model using incoming and historical data. This processing type
requires that the model state is retained. Further, access to additional data
is often required.

The main difference between streaming applications with traditional, data-
intensive batch applications is that streaming data sources are unbounded.
While this impacts some aspects of an applications, such as the runtime and
the potentially need to carefully reason about ordering and time constraints,
other factors remain the same, e.\,g., the computational complexity of the
processing algorithms. In the following, we utilize the following sub-set of
properties to characterize streaming applications:

\textbf{Data Source and Transfer:} describe the location of the data
source in relation to the stream processing application. The data source can be
external (e.\,g., an experimental instrument) or internal to the application
(e.\,g., the coupling of a simulation and analysis application on the same
resource). Output data is typically written to disk or transferred via a
networking interface. Message brokers can serve as intermediate decoupling
production and consumption.

\textbf{Latency} is defined as the time between arrival of new data and
its processing.

\textbf{Throughput} describes the capacity of the streaming system,
i.\,e. the rate at which the incoming data is processed.

\textbf{Lifetime:} Streaming applications operate on unbounded
data streams. The lifetime of a streaming application is often dependent on the
data source. In most cases it is not infinite and limited to e.g., the
	simulation or experiment runtime.
	
\textbf{Time/Order Constraints} defines the importance of order while
processing events.

\textbf{Dynamism:} is variance of data rates and
processing complexity observed during the lifetime of a streaming application.

\textbf{Processing:} This characteristics describes the
complexity of data processing that occurs on the incoming data. It depends e.\,g. on the amount of data being processed (window size, historic data) and the algorithmic complexity.

\subsection{Streaming Application Examples}
\label{sec:lightsource}

\alnote{Reviewer: the “Streaming in Scientific Application” section would be better integrated with the following “Application Characteristics” section. Use them as examples of the different types of applications.}

In the following we utilize the defined streaming application characteristics
to analyze two example use cases: (i) a generic streaming analytics application
(Type 1 or 2), and a more specific use case (ii) light sources analytics (Type
1). Table~\ref{tab:streaming_app_characteristics} summarizes different
characteristics of these applications.

\subsubsection{Streaming Analytics}

Use cases, such as Internet-of-Things, Internet/Mobile clickstreams, urban
sensor networks, co-analysis of simulation data, demand the timely processing
of data feeds using different forms of analysis~\cite{streaming2015,
Gannon-2016}. For example, an increasing number of scientific applications
require streaming capabilities: cosmology simulations require increasing
amounts of data analytics to digest simulation data, environmental simulation
require the integration of remote sensing capabilities, etc. Depending on the
nature of the data source, this type of application can be classified as type 1
or 2 application. The number of type 3 application is still comparable low. 
This can be attributed to the lack of sufficient middleware to support such 
complex architectures.

While the general problem architecture of data analytics and machine learning
are similar to those of batch application, there are some subtle differences:
typically the amount of data processed at a time is small compared to batch
workloads. While the problem architecture of many machine learning algorithms
remains the same, different techniques for updating the model using the new
batch of data are used (e.g., averaging using a decay factor).

\subsubsection{Light Source Sciences}

X-Ray Free Electron Laser (XFEL) are a class of scientific instruments that
have become instrumental for understanding fundamental processes in domains
such as physics, chemistry and biology~\cite{lightsource-grand-challenge,
Hand2009}. Such light sources can reveal the structural properties of proteins,
molecular and other compounds down to the atomic levels. The light source emits
hundreds to thousands of x-ray pulses per second. Each pulse produces an image
of the diffraction pattern as results. These images can then be combined and
reconstructed into a 3-D model of the compound serving as the basis for a later
analysis. Light sources can be used to exactly observe what is happening during
chemical reactions and natural processes, such protein folding.

Example for light sources are the Linac Coherent Light Source
(LCLS)~\cite{linac} at SLAC, the National Synchroton Light Source II (NSLS
II)~\cite{nsls} at Brookhaven, and the European XFEL light
sources~\cite{european_xfel}. LCLS-I averages a throughput of 0.1-1 GB/sec with
peaks at 5 GB/s utilizing 5 PB of storage and up to 50 TFlops
processing~\cite{lcls_data}. The European XFEL produces 10-15 GB/sec per
detector~\cite{european_xfel}. In the future even higher data rates are
expected: LCLS-II is estimated to produce data at a rate of more than 20
GB/sec.  In the following, we focus on NSLS-II.

NSLS-II consists of 29 operational beamlines. Thirty more beamlines
are in development. Each beamline has different data characteristics, therefore
the need for developing management tools that acquires the data from the
beamlines and analyzes them is evident. As the beamtimes available to the user
are typically short, it essential that processing and analysis can be conducted
in a timely manner. Thus, streaming data analysis is an important capability to
optimize the process. This ensures that scientists can adjust the settings on
the beamline and optimize their experiment.

The Complex Materials Scattering (CMS) beamline is an NSLS-II beamline,
which generates 8\,MB images at a rate of 10 images/minute. While this
production data rate is not very high, a single CMS experiment generates more
than 17,000 images a day, equivalent to $\sim$140 GB of data. It is required
that this data be processed within 6 hours, to prepare for the experiments the
following day. The Coherent Hard X-ray (CHX) beamline~\cite{CHX} is dedicated
to studies of nanometer-scale dynamics using X-ray photon correlation
spectroscopy can produce data at much higher rates of
$\sim$4.5GB/s~\cite{nslsTalk}.

Light source applications are a Type 1 application. In most cases, the
instrument is co-located with some compute resources. However, scientists often
rely on additional compute resource and also may need to integrate data from several instruments. Thus, the ability to manage geographically distributed resources is important. Currently, data analysis is often decoupled from the
experiments. With increased sophistication of the instruments, the demand for
steering capabilities will grow evolving this type of application toward Type 
3.

The processing pipeline for light source data comprises of three
stages: pre-processing, reconstruction and analysis~\cite{tomopy}.
Pre-processing can includes e.\,g. normalization of the data, filtering and the
correction of errors. Various reconstruction with different properties, e.\,g.
computational requirements and quality of the output, exist:
GridRec~\cite{gridrec} is based on a Fast-Fourier transformation and is less
computational intensive and thus, fast. Iterative methods can provide a better
fidelity. An example of an iterative method is Maximum likelihood expectation
maximization (ML-EM) reconstruction~\cite{mlem}. A broad set of analytics
methods can be applied  to the reconstructed image, e.\,g. image
segmentation and deep learning methods. For the CMS experiment, 
simple statistical algorithms, such as the computation of a circular average 
and peak finding is used.

\subsubsection{Discussion}

\begin{table}[t]
	\scriptsize
	\centering
\begin{tabular}{|p{1.3cm}|p{3.3cm}|p{3.3cm}|} \hline
\textbf{} 		&\textbf{Streaming Analytics: K-Means} 		&\textbf{Light Source}\\ \hline %

Data Source    &external or internal &external  \\ \hline

Latency 		&medium/high latencies 
				&medium latencies		\\ \hline

Throughput 		&medium  	&high \\ \hline
Duration			 		&data source runtime  	&experiment runtime \\ \hline

Time/Order  		&not important  	&not important \\ \hline

Dynamism 			&varying data rate  	&varying data rate \\ \hline

Processing    		    &\textbf{Model score:} Assign data to centroids/class $O(num\_points \cdot num\_clusters)$. \textbf{Model update:} Update centroids with in-coming mini-batch of data. \textbf{Model size:} small (O(number clusters))
	 			        &\textbf{Reconstruction:} Reconstruction techniques with different complexities (GridRec, ML-EM). \textbf{Analysis:} data analysis techniques, such as peak finding, image processing models utilizing GPUs. \\
						 \hline
\end{tabular}
\caption{Streaming Application Properties\label{tab:streaming_app_characteristics}\upp\upp\upp\upp\upp\upp}
\end{table}

The requirements of streaming applications vary: For use cases involving
physical instruments with potential steering requirement, e.\,g., X-Ray Free
Electron Laser, both latency and throughput are important. Other use cases
e.\,g.\ the coupling of simulation and analysis have less demanding latency and
throughput requirements. The lifetime of scientific streaming applications is
often coupled to the lifetime of the data source. Time and message ordering is
in contrast to transactional enterprise applications not important for many
scientific applications. With respect to the data transfer and processing
requirements, the need to support different frameworks in a plug-able and
interoperable way is apparent.

Another important difference is that streaming applications are typically
runtime constrained, i.\,e. they must process the incoming data at a certain
rate to keep the system balanced. Thus, a good understanding of application
characteristics is even more critical for streaming applications. Minor changes
in the data rates, the processing approach (e.\,g. change of the processing
window, sampling approaches or the need to process additional historic data or
available resources) can lead to imbalance and a dysfunctional system. Thus, the
ability the dynamically allocate additional resources to balance the system is
critical. We use the characteristics identified in this section to design the
Streaming Mini-Apps that aids the evaluation of complex streaming
systems (see section~\ref{sec:miniapps}).

\section{Pilot-Streaming: Abstractions, Capabilities and Implementation}
\label{sec:arch_impl}

Pilot-Streaming addresses the identified challenges and gaps related to
deploying and managing streaming frameworks and applications on HPC
infrastructure. Pilot-Streaming makes two key contributions: (i) it defines a
high-level abstractions that provide sufficient flexibility to the application
while supporting the resource management and performance needs of streaming
applications are essential, and (ii) the reference implementation supports
different stream processing and brokering frameworks on HPC resources in a
plug-able and extensible way.

Pilot-Streaming provides a well-defined abstraction, i.\,e. a simplified and
well-defined model that emphasizes some of the system's details or properties
while suppressing other~\cite{shaw1984}, for managing HPC resources using
\pilot-Jobs and deploy streaming frameworks on these. The Pilot-Stream
abstraction is based on the Pilot-Job abstraction. A Pilot-Job is a system that
generalizes the concept of a placeholder job to provide multi-level scheduling
to allow application-level control over the system scheduler via a scheduling
overlay~\cite{pstar12}. \pilotjobs have been proven to provide efficient
mechanisms for managing data and compute across different, possibly distributed
resources. The Pilot-Abstraction is heavily used by many HPC application for
efficiently implementing task-level parallelism, but also advanced execution
modes, such as processing of DAG-based task graphs. Examples for using the
Pilot-Abstraction are molecular dynamics simulations~\cite{balasubramanian2016extasy} and
high energy application~\cite{turilli2017comprehensive}. Further, we have
explored the applicability of the
Pilot-Abstraction~\cite{saga_bigjob_condor_cloud} to data-intensive
applications on HPC and Hadoop environments~\cite{luckow2015pilot,
luckow2016hadoop}. 

\revThreeNote{The fault tolerant aspect is not really discussed and unfortunately hardware faults can be added as an interesting topic when dealing with resource allocation management but not sure this paper is intended to deal with such topic.}\alnote{added sentence. No space for in-depth discussion}

The \pilot-Streaming reference implementation allows the management and
deployment of different message brokers and stream processing frameworks,
currently Spark, Dask and Kafka, as well as its ability to serve as unified
access layer to run tasks across these in an interoperable way. Further, these
frameworks can be deployed side-by-side on the same or different
distributed resources a capabilities which is critical for many streaming
pipelines. The framework is designed in extensible way and can easily be
extended to support Flink, Heron and other stream processing frameworks.
Another key capability is the ability to dynamically scale these frameworks by
adding resources. This is essential to deal with varying data rates and compute
requirements. Further, framework continuously monitors the applications and
thus, provides an enhanced level of fault tolerance, which is essential as
stream applications typically run longer than batch jobs. We continue with a 
discussion of the Pilot-Streaming abstraction in section~\ref{sec:abstraction} 
and the reference implementation in section~\ref{sec:ref_impl}.

\subsection{Pilot-Abstractions and Capabilities}
\label{sec:abstraction}

In this section, we describe the provided abstraction from developer point of
view. The abstraction is based on the Pilot-abstraction, which provides two key abstractions: a Pilot
represents a placeholder job that encapsulates a defined set of user-requested
resources. Compute-Units are self-contained pieces of work, also referred to as
tasks, that are executed on these resources. Pilot-Streaming utilizes
multi-level scheduling and can manage Compute-Units in a framework-agnostic
way. For this purpose, Pilot-Streaming interfaces with the schedulers of the
different frameworks, e.\,g. the Spark scheduler, which then manage the
further execution of the Compute-Units. The key features of
Pilot-Streaming are:

\textbf{Unified and Programmatic Resource Management:} The 
	\pilot-Abstraction provides a unified resource management abstraction to 
	manage streaming frameworks for processing and message brokering on HPC 
	environments. It allows the orchestration of compute and data across 
	different frameworks.

\textbf{Streaming Data Sources:} While our previous work 
	focused on integration static datasets and compute 
	units managed by Pilot-Jobs~\cite{luckow2015pilot}, Pilot-Streaming extends this 
	ability to streaming data sources, such as Kafka topics.

\textbf{Interoperable Streaming Data Processing:} For the processing 
	of streaming data applications can utilize the Pilot-API for defining 
	\computeunits. 	\computeunits can either rely on native HPC libraries and 
	applications or can integrate with stream processing frameworks, such as 
	Spark-Streaming.  This enables applications to utilize the different 
	capabilities of these frameworks in a unified way.
	
\textbf{Extensibility and Scalability:} Pilot-Streaming is extensible 
	and can easily be extended to additional message brokers and streaming 
	frameworks. It is architected to scale to large (potentially distributed) 
	machines both at deploy and runtime.

The framework exposes two interfaces: (i) a command-line interface and
(ii) the Pilot-API for programmatic access. The API is based on a well-defined
conceptual model for Pilot-Jobs~\cite{pstar12}.  The Pilot-API allows
reasoning about resources and performance trade-off associated with streaming
applications. It provides the means necessary to tune and optimize application
execution by adding/removing resources at runtime.
Listing~\ref{lst:pilot_init} shows the initialization of a Pilot-managed Spark
cluster. The user simply provides a pilot compute description object, which is
a simple key/value based dictionary.

\revThreeNote{How does Pilot-Streaming manages dynamically the resources needed by stream engines? Authors discuss the requirements to dynamically adjust the number of resources (auto-scale) but do not explain if this is possible with their architecture and how.
}\alnote{describe below. Listing not possible due to space reasons}

\begin{lstlisting}[language=python, basicstyle=\ttfamily\scriptsize, breaklines,  captionpos=b, caption=Pilot-Streaming: Creation of Spark Cluster, label=lst:pilot_init]
from pilot.streaming.manager import PilotComputeService
spark_pilot_description1 = {
    "service_url":
           "slurm+ssh://login1.wrangler.tacc.utexas.edu",
    "number_cores": 48,
    "type":"spark"
}
pilot1 = PilotComputeService.create_pilot(spark_pilot_description)
 \end{lstlisting}

A key capability of Pilot-Streaming is the ability to dynamically add/remove
resources to the streaming cluster by just referencing a parent cluster in the
Pilot-Description.  If the resources are not needed anymore, the pilot
can be stopped and the cluster will automatically resize. This capability not
only allows application to respond to varying resource needs, but also provides
the ability to work around maximum job size limitations imposed by many
resource providers.

Pilot-Streaming provides several hooks to integrate with the managed streaming
frameworks. It supports custom configurations, which can be provided in their
framework native form (e.g., spark-env format etc.) and can easily be managed on
per machine basis. This ensures that machine-specific aspects, e.g., amount
of memory, the usage SSD and parallel filesystems, network configurations, can
optimally be considered.

Pilot-Streaming supports interoperability on several levels. The API provides a
unified way to express stream computations agnostic to specific framework.
Listing~\ref{lst:interopcu} illustrates how to execute a Python function can be
executed as a \computeunit in a interoperable way. This is suitable for simple
stream-processing tasks, such as tasks that can be expressed as map-only job.
Using the unified API, functions can easily be run across frameworks, e.\,g. to
utilize advanced, framework-specific capabilities, such as parallel processing,
windowing or ordering guarantees. For more complex tasks, the API provides the
ability to access the native API of each framework allowing the implementation
of complex processing DAGs.

\begin{lstlisting}[basicstyle=\ttfamily\scriptsize, captionpos=b, caption=Pilot-Streaming: Interoperable Compute Unit, label=lst:interopcu]
def compute(x):	return x*x
compute_unit = pilot.submit(compute, 2)
compute_unit.wait()
\end{lstlisting}

Listing~\ref{lst:nativecu} illustrates how the Context-API provides the ability
to interface with the native Python APIs from these frameworks. The context
object exposes the native client application, i.\,e., the Spark Context, Dask
Client or Kafka Client object. Having obtained the context object, the user can
then utilize the native API,  e.g., the Spark RDD, DataFrame and Structured
Streaming API.

\begin{minipage}{\linewidth}
\begin{lstlisting}[basicstyle=\ttfamily\scriptsize, captionpos=b, caption=Pilot-Streaming: Native Spark API Integration, label=lst:nativecu]
sc = spark_pilot1.get_context()
rdd = sc.parallelize([1,2,3])
rdd.map(lambda x: x*x).collect()
\end{lstlisting}
\end{minipage}

\subsection{Reference Implementation: Architecture and Interactions}
\label{sec:ref_impl}

\begin{figure}[t]
  \centering
 \includegraphics[width=.45\textwidth]{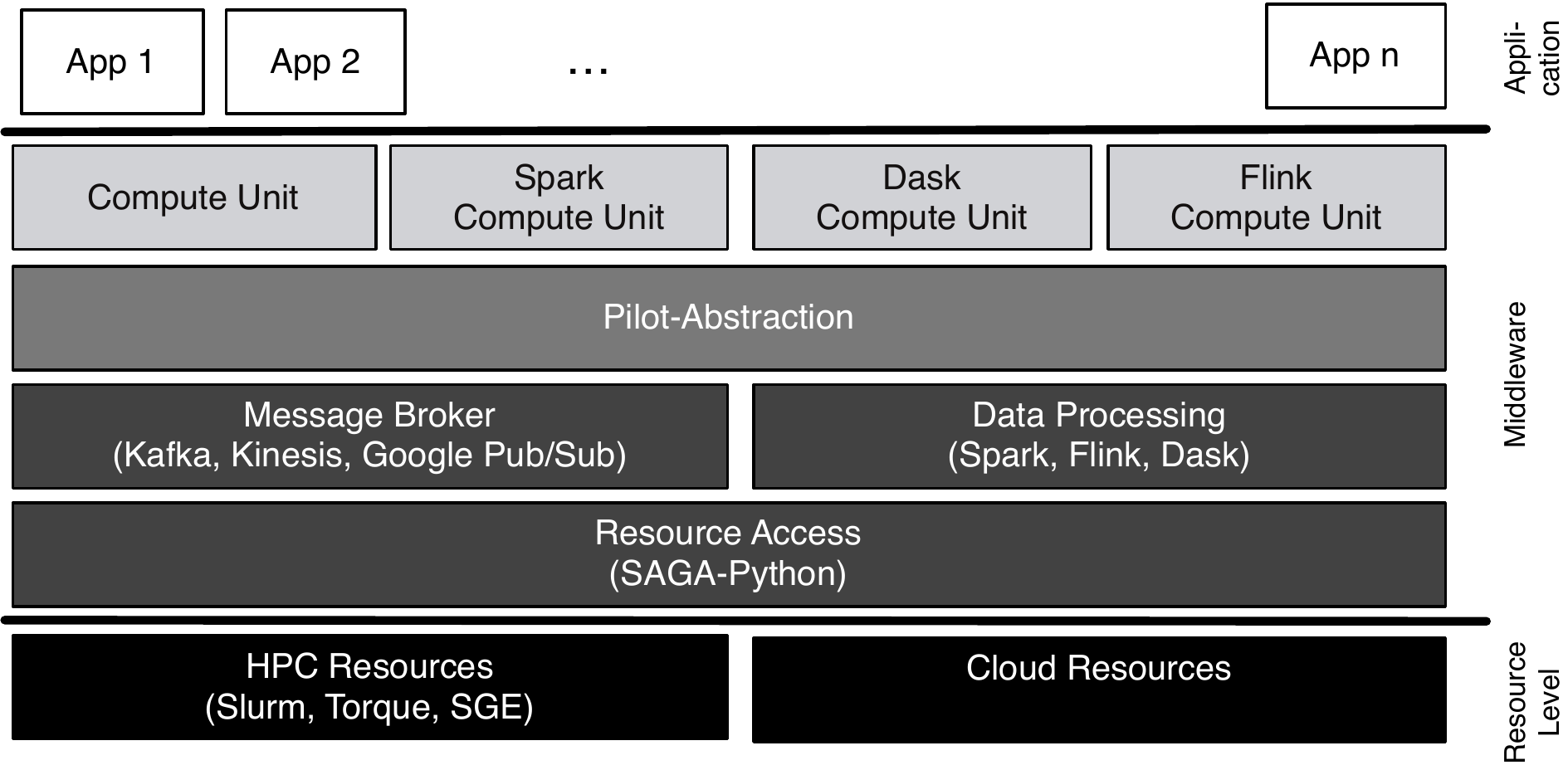}
  \caption{\textbf{Pilot-Streaming Architecture:} Pilot-Streaming allows the management  of message brokers and stream processing frameworks on HPC.\upp\upp}
  \label{fig:figures_streaming_hpc}
\end{figure}

Figure~\ref{fig:figures_streaming_hpc} illustrates the high-level architecture
of Pilot-Streaming. Pilot-Streaming provides a unified access to both HPC and
cloud infrastructure. For resource access we utilize the SAGA Job
API~\cite{merzky2015saga}, a lightweight, standards-based abstraction to resource
management systems, such as SLURM, SGE and PBS/Torque. The framework provides
two key capabilities: the management of message broker on HPC and the
management of distributed data processing Engines on HPC. These two
capabilities are encapsulated in the message broker and data processing module.
The interface to the framework is the Pilot-Abstraction~\cite{pstar12}, a
proven API for supporting dynamic resource management on top of HPC machines.
The application logic is expressed using so-called Compute-Unit, which can be
executed in either (i) a task-parallel processing engine, such as Pilot-Jobs
(e.\,g., RADICAL-Pilot~\cite{rp-jsspp18},
BigJob~\cite{saga_bigjob_condor_cloud} or Dask), or (ii) a streaming
framework, such as Spark Streaming. Case (i) typically requires the
manual implementation of some capabilities, e.\,g.\ the continuous polling of
data. In case (ii) the developer can rely on the streaming
framework for implementing windowing. Both scenarios have
trade-offs: while scenario (i) allows the interoperable execution of CUs across
frameworks, scenario (ii) is often faster to implement. Pilot-Streaming
supports both cases.

\begin{figure}[t]
  \centering    \includegraphics[width=.4\textwidth]{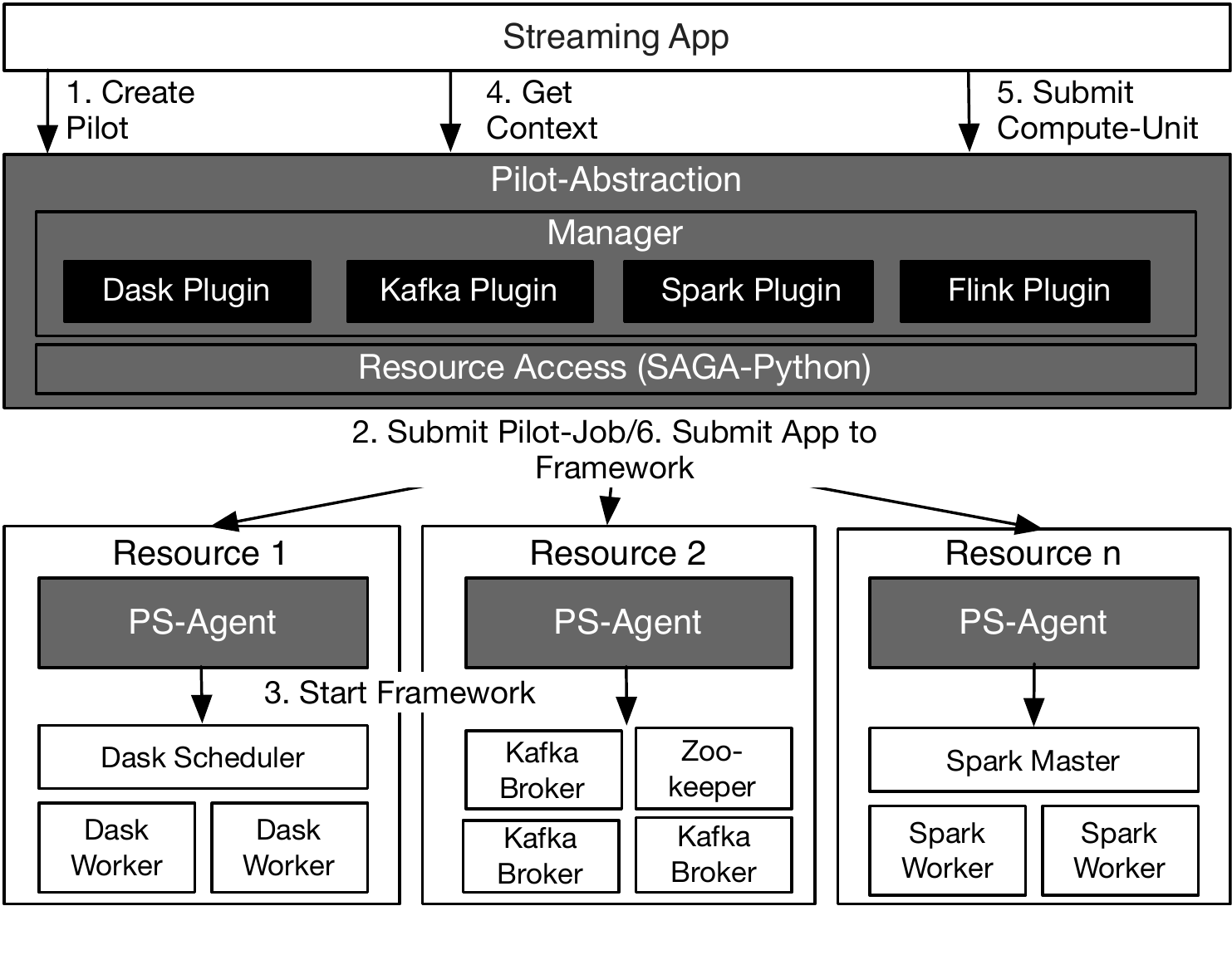}
  \caption{\textbf{Pilot-Streaming Interaction Diagram:} The figure shows the control flow used by Pilot-Streaming to manage frameworks and applications.\upp\upp}
  \label{fig:figures_pilotstreaming-interaction}
\end{figure}

Figure~\ref{fig:figures_pilotstreaming-interaction} shows the interaction diagram for Pilot-Streaming.
In the first step the application requests the setup of Spark, Dask or Kafka
cluster using a Pilot-Description as specification. Then the Pilot-Manager 
initiates a new \pilotjob, a placeholder job for the data processing or message
broker cluster, via the local resource manager. The component running on
resource is referred to as Pilot-Streaming-Agent (PS-Agent). After the job and
framework has been initialized, the application can start to submit
\computeunits or initiative interactions with the native framework APIs via the
context object.
\pilot-Streaming is an extensible framework allowing the simple addition of new
streaming data sources and processing frameworks. By encapsulating important
components of streaming applications into a well-defined component and API,
different underlying frameworks can be used supporting a wide variety of
application characteristics. It utilizes the SAGA-Python~\cite{saga-python-pd}
implementation to provision and manage resources on HPC machines. 

The streaming frameworks specifics are encapsulated in a plugin. A framework
plugin comprises a simple service provider
interface (SPI) and a bootstrap script executed on the resource. As depicted in
Listing~\ref{lst:plugin}, the interface has  six functions, e.\,g., to
start/extend a cluster, to retrieve cluster information, such as state and
connection details.

\begin{lstlisting}[language=python, basicstyle=\ttfamily\scriptsize, breaklines, captionpos=b, caption=Pilot-Streaming Plugin Interface, label=lst:plugin]
class ManagerPlugin():
    def __init__(self, pilot_compute_description)
    def submit_job(self)
    def wait(self)
    def extend(self)
    def get_context(self, configuration)
    def get_config_data(self)
\end{lstlisting}

\begin{table}[t]
	\centering
	\scriptsize
	\begin{tabular}{|p{1.5cm}|p{6.0cm}|}\hline
	Data Source & Adhoc deployment of broker and processing close to data \\ \hline
	Latency & Framework selection, co-location of data/compute resource \\\hline
	Throughput & Framework selection, optimization of resource configuration to data rate \\\hline
	Fault Tolerance & Monitoring of Jobs through Pilot-Job Management and Agent \\\hline
	Time/Ordering  & Orderting, Windowing mechanism of underlying framework  \\\hline
	Dynamism & Add/Remove resources at runtime via Pilot-Job Mechanism \\\hline
	\end{tabular}
	\caption{Streaming Challenges Addressed by Pilot-Streaming\label{tab:ps} \upp\upp\upp\upp\upp}
\end{table}

\emph{Discussion:}
Data and streaming applications are more heterogeneous and complex than
compute-centric HPC applications. Pilot-Streaming allows the usage of
different message brokers and data processing engines in an interoperable way on
HPC infrastructures. Table~\ref{tab:ps} summarizes how Pilot-Streaming addresses the requirements of streaming applications.

Pilot-Streaming removes the need for application developers
to deal with low-level infrastructure, such as resource management systems.  Running Spark, Kafka  and Dask clusters
across a flexible number of Pilot-Jobs provides the ability to dynamically
adjust resources  during runtime. Further, the framework provides a
common abstraction to execute compute tasks and integrate these with streaming
data. It supports the interoperable execution of these CU across different
frameworks. In addition, Pilot-Streaming provides the ability to also utilize
the higher-level APIs provided by the frameworks. Currently, Pilot-Streaming
supports Kafka, Spark, and Dask. It can  be extended via a
well-documented plugin-interface. Pilot-Streaming is open-source, maintained by
an active developer community and available on Github~\cite{pilot-streaming}.

\alnote{Add reasoning about Multi-level resource management, many optimization
opportunities, e.g., ratio number workers/to resource course}

\upp
\section{Streaming Mini-Apps}
\label{sec:miniapps}
\upp

\revThreeNote{The authors develop and describe a synthetic simulation/benchmark framework (i.\,e., Streaming Mini-Apps) used to develop streaming applications (e.\,g., K-Means, Light-source) built on top of the Pilot-Streaming. It is not clear why another API is needed instead of leveraging existing APIs (Kafka, Spark/Flink).
}

Developing streaming application pipelines is a complex task as it requires
multiple parts: data source, broker and processing component. Every one of these
components typically relies on different programming and middleware systems
making it highly complex to develop such pipelines. During development process
the real data source is often not available. Often developers have to rely on a
static dataset, which results in significant efforts for setting setup a real
test and development environment that is capable of mimicking non-bounded
datasets as well as non-functional requirements, such as different data rates,
message sizes, serialization formats and processing algorithms. If available,
real applications are often not as parameterizable and tunable to characterize
and optimize application, middleware and infrastructure configurations.

\begin{figure}[t]
  \centering
    \includegraphics[width=.35\textwidth]{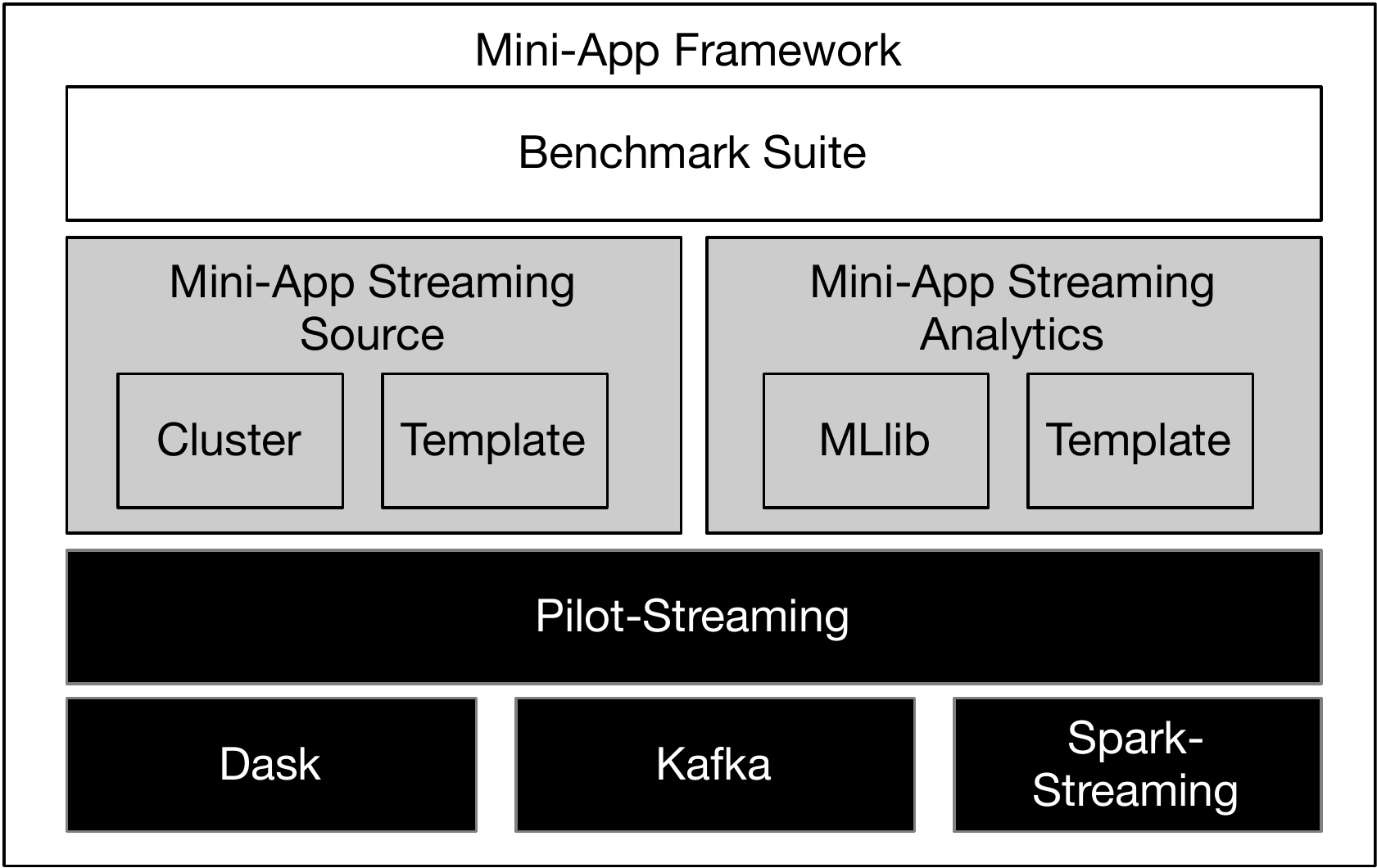}
  \caption{Streaming Mini-Apps: The framework is based on Pilot-Streaming and provides two components: the \emph{MASS (MiniApp for
Stream Source)} emulates different  streaming data sources and the \emph{MASA (MiniApp for Streaming Analysis)} provides different synthetic processing workloads.\upp\upp\upp\upp}
  \label{fig:figures_miniapp}
\end{figure}

The Streaming Mini-Apps~\cite{miniapps} addresses these challenges.
Figure~\ref{fig:figures_miniapp} shows the architecture of the 
framework. The framework is based on Pilot-Streaming, which provides the
ability to rapidly allocate different size of cluster environments. The core of
the framework consists of two main components: (i) the \emph{MASS (Mini-App for
Stream Source)} can emulate a streaming data source, which can be tuned to
produce streams with different characteristics: data rates, messages size. (ii)
the \emph{MASA (Mini-App for Streaming Analysis)} provides a framework for
evaluating different forms of stream data processing.

The MASS app includes a pluggable data production functions. The current
framework provides two types of functions: A cluster source generates random
data points following certain structures, e.\,g., for evaluation of streaming
cluster analysis algorithms. The second type: template produces an unbounded
stream based on a static template dataset. Data rates, message sizes etc. can
be controlled via simple configuration options. Using these two base data
source the majority of streaming applications can be emulated. For example,
KMeans or other cluster algorithms for detecting outliers in data streams can
be developed and tested with the cluster source. The template algorithms is
great for migrating batch workloads to streaming. It can be used to emulate
important application, such as light sources.

Similarly, the MASA app enables the user utilize machine learning algorithms
from MLlib~\cite{Meng:2016:MML:2946645.2946679} or to provide custom data
processing functions. Currently, it is based on Spark Streaming, but the
framework can easily ported to other streaming frameworks as it is based on
Pilot-Streaming. The processing function is data-parallel by nature. The
machine learning algorithms provided by MLlib are capable of utilizing
distributed resources supporting both data and model parallelism.
In particular, we provide pre-configured support for KMeans
clustering~\cite{spark-streaming-clustering} and for reconstructing light
source data. The K-Means algorithm has a complexity of $O(cn)$ where $c$ is the
number of cluster centroids and $n$ is the number of data points. The light
source reconstructing algorithm is based on Tomopy~\cite{tomopy}, a framework
that is commonly used for pre-processing raw light source data, e.\,g., image
reconstructions, and for further analysis. Different reconstruction algorithms 
are supported by the Mini-Apps, e.\,g.,  GridRec~\cite{gridrec}
and ML-EM~\cite{mlem}.

In summary, the Streaming Mini-Apps provide optimal customizability
with the ability to plug in custom data production and processing functions and control various configuration parameters, such as data rates, message sizes,
etc. The framework provides comprehensive performance analysis options, e.\,g.
it includes standard profiling probes that enables to measure common metrics,
such as production and consumption rate allowing the benchmark of application
and streaming middleware components making it easy to understand performance
bottlenecks as well as the impact of changes. This is an essential capability
to develop, test and tune streaming pipelines under complex, real world loads.
In particular components like the message broker are difficult to analysis as
the write/read load can vary significantly depending on the number consumers
and producers. Further, the Mini-Apps allow for easy reproducibility
of such experiments. The Streaming Mini-Apps provide a powerful tool to
develop, optimize applications, and empirically evaluate streaming frameworks
and infrastructure. In contrast to other approaches~\cite{7530010}, the
streaming mini app framework focuses on data-related characteristics, in
particular the need to produce, transport and process data at different rates.
In addition, the framework can emulate the application characteristics of
K-Means application.

\section{Experiment and Evaluation}\label{sec:exp}

The aim of this section is to investigate different infrastructure
configuration with respect to their ability to fulfill defined application
requirements in terms of latency and throughput. For this purpose, we use
the Mini-Apps to simulate different data production and processing 
characteristics.  All experiments are conducted on Wrangler, an XSEDE machine 
designed for data-intensive processing. Each Wrangler nodes has 128\,GB of 
memory and 24 cores.

\subsection{Startup Overhead}

\begin{figure}[t]
  \centering
\includegraphics[width=.4\textwidth]{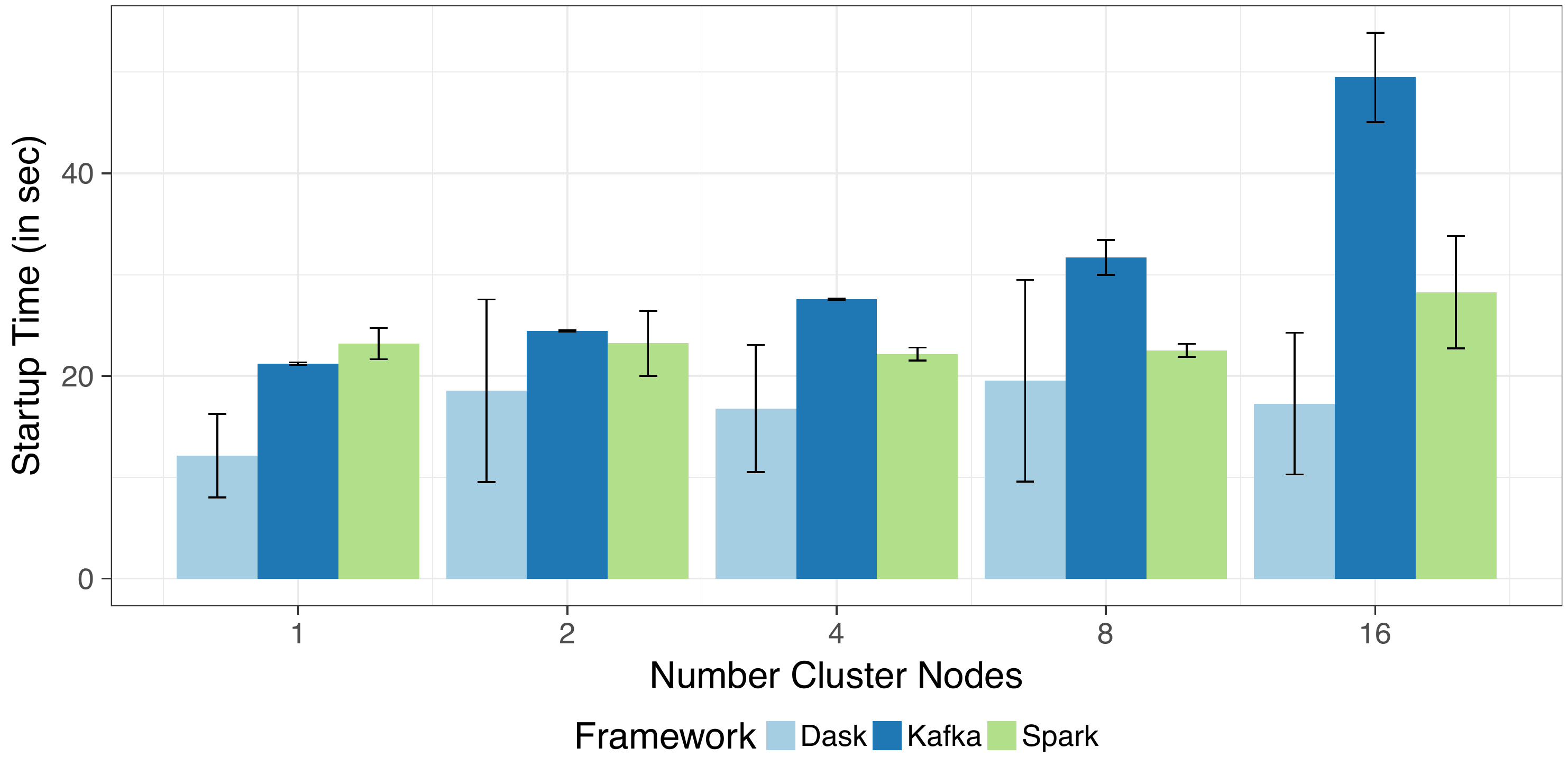}
  \caption{\textbf{Kafka, Spark, and Dask Startup Time 
  on Wrangler:} Kafka start involves the startup of both Zookeeper and the
  Kafka brokers and thus, is most of the times longer than Spark. Dask has the 
  shortest startup times. For Kafka, the startup time increase with the number 
  of nodes. The Spark and Dask startup times did not significantly change for 
  larger clusters.\upp\upp\upp\upp}
  \label{fig:experiments_startup_kafka_spark_startup}
\end{figure}

There are two main steps for setting up Spark and Kafka on HPC: (i) Running the
batch job that sets up the Kafka/Spark cluster and (ii) initiating an actual
session with the broker respectively starting a Spark job by initializing a
Spark session. Figure~\ref{fig:experiments_startup_kafka_spark_startup}
compares the startup times for different size Kafka, Spark and Dask clusters. 
The startup time for Kafka increase significantly with the number of nodes 
indicating that some optimizations are necessary for larger clusters. Spark and 
Dask utilize parallelism to startup the cluster and thus, show no significant 
increase.

The measured startup times are short compared to the overall runtime
of streaming application. In particular, considering the benefits of
Pilot-Streaming: improved isolation of application components, the ability to
independently scale parts of the streaming pipeline to the application needs,
better diagnose-ability, debug-ability and predictability of the application,
this is an acceptable overhead.

\subsection{Producer Throughput}

In this section, we analyze the performance for publishing data into the Kafka
system using the MASS app. The produces batches of random 3-D points, which are
serialized to a string and pushed to Kafka using PyKafka~\cite{pykafka}. We
utilize different data source types: (i) KMeans: every message consists of
5,000\, randomly generated double precision points. The average serialized size
of message is 0.32\,MB; (ii) Lightsource Micro-Tomography (Light-MT): every
message consists of raw input dataset in the APS data format and an average
encoded message size of 2\,MB. (iii) Lightsource CMS (Light-CMS): every message
consists of one image generated from the CMS Beamline. The size of each image
is 8\,MB (HDF5) and 18\,MB (serialized). The scenarios were chosen to
demonstrate the variety characteristics with respect to number messages and
message sizes streaming application can exhibit.

\alnote{@George: Do we have some data for this? Contention can be due to
various factors: network, filesystem, memory.... In all three scenarios that
we are investigating we observe that the throughput is increasing and after
the point that it reaches the peak it starts reducing, even though the number
of producers is increasing. That is happening because of the I/O contention on
the filesystem. More specifically, according to the  gpfs~\cite{gpfs_paper}
paper, it has a performance limitation when the producer, in this case Kafka,
is sending messages faster than the gpfs can save the data to disks, it
enables an backoff protocol in order to handle the incoming messages.The
application node is delayed y, after it retries, then 2y, 4y, 8y etc. As a
result the producer throughput  is reduced.  Adding extra broker to the system
resolves  the I/O contention issue. }

\begin{figure}[t]
  \centering
\includegraphics[width=.49\textwidth]{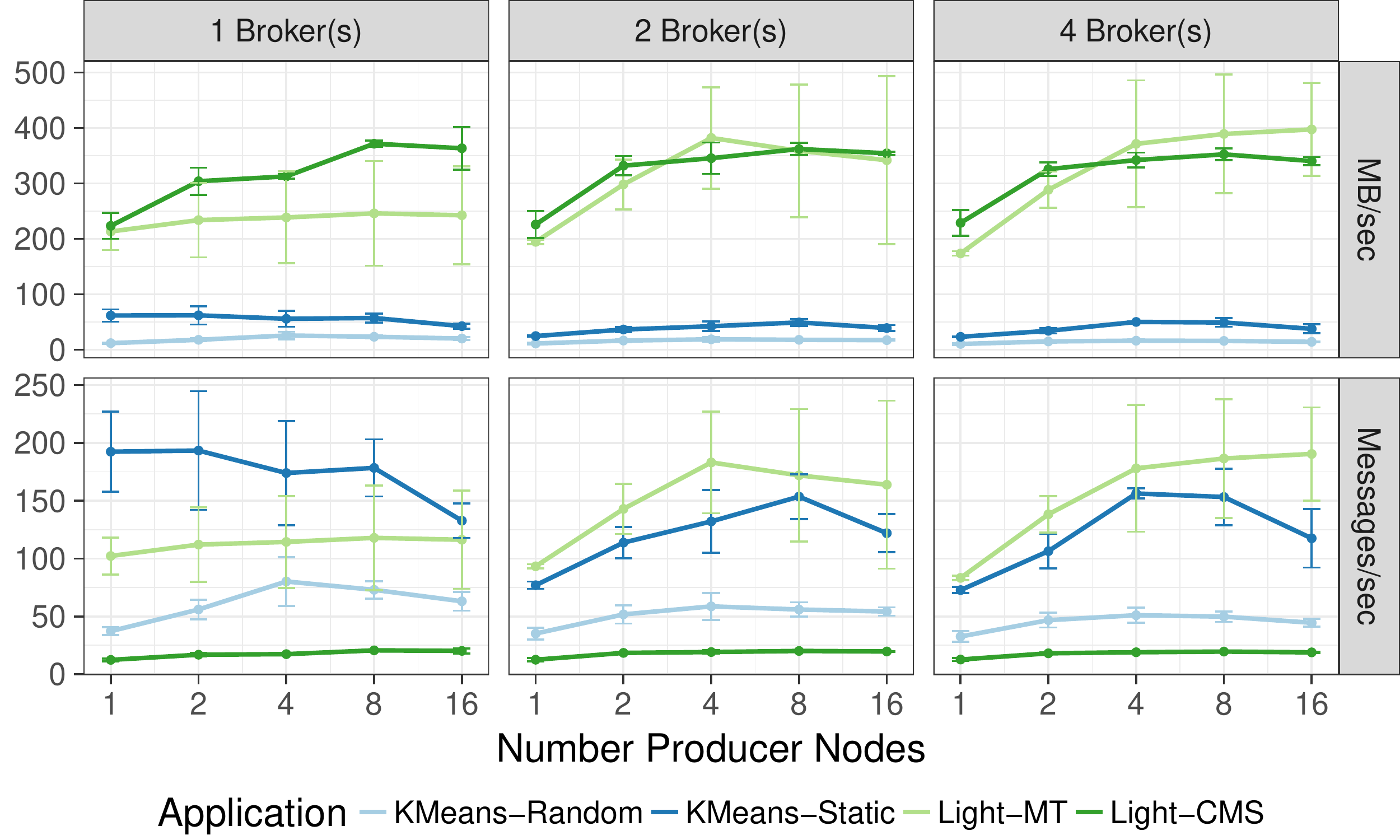}
  \caption{\textbf{MASS Producer Throughput for Different Data Sources Types 
  and Resource Configurations:} We utilize up to 16 producer nodes with 8   
  processes/node and 4 Kafka nodes. The achievable throughput depends on the 
  message size: KMeans: 0.3\,MB, Light-MT: 2\,MB and for Light-CMS: 18\,MB.\upp\upp\upp }
  \label{fig:experiments_kmeans_light}
\end{figure}

We investigate the throughput and its relationship to different MASS types and
configurations as well as to different Kafka broker cluster sizes. For  the
experiment, we utilize different resource configuration parameters
determined in a set of micro-experiments: the number partitions is fixed at
12 per node. On every producer node, we run 8 producer processes in Dask. While
each node possesses 24 cores, the performance per node deteriorated
drastically when using more producers/node due to network and I/O bottlenecks.
We evaluate four scenarios: KMeans-Random, KMeans-Static, Light-MT and 
Light-CMS. The KMeans-Random scenario uses the cluster MASS plugin to generate 
points randomly distributed around a defined number of centroids. Kmeans-Static 
and both light scenarios use a static message at a configured rate.

Figure~\ref{fig:experiments_kmeans_light} shows the results. The KMeans-Random
configuration is bottlenecked by the random number generator. Thus, the
KMeans-Static setup has on average a 1.6x higher throughput than KMeans-Random.
The light scenarios show a significant MB/sec throughput mainly due to larger
message sizes: Light-CMS uses a much larger message size (18\,MB) compared to 
Light-MT (2\,MB), thus the throughput is in many cases higher for Light-CMS 
than for Light-MT. As expected, the message throughput is lower for 
Light-CMS due to the larger message sizes. Both the message throughput and high 
variance in the measured bandwidth indicate that the performance is network 
bound. Also, it must be noted that the network is a shared resource and 
external factors likely lead to the high variance in the measured 
bandwidths for Light-CMS and Light-MT. The usage of more brokers does not 
improve the performance in all scenario due to the overhead associated with 
accessing a multi-node Kafka cluster, e.\,g. concurrent connections and 
partitioning overhead. A multi-node Kafka cluster is particular advantageous 
when a larger number of medium-sized messages need to be handled, such as for
Light-MT.

\alnote{Where do we see the latency spikes in the data: Larger messages are causing more longer flushes to the filesystem, which lead to more and longer latency spikes, which leads to reduced throughput.}

\upp
\subsection{Processing Throughput}
\upp

\revTwoNote{Reviewer: 
It is maybe obvious for expert but some results are not discussed, for example the fact that for 1 Broker case the throughput in MB/sec for Light-CMS is above Light-LT but for 2 and 4 Brokers they are very close to each other, for KMeans-Static and Light-MT concerning the Messages/sec why the KMeans-Static is better with 1 Broker and worst with more compared to Light-MT. 
In the sentence “The 1 broker scenario shows a limited throughput for all scenarios” is not true for all scenarios (for KMeans-Static it is wrong for example).
All these approximations or mistakes (if true ones) lead to confusions and/or misunderstanding of the results.
}

We use the MASA Mini-App to investigate the throughput of three different
processing algorithms: a streaming KMeans that trains a model with
10 centroids and makes a prediction on the incoming data, and two light source
reconstruction algorithms: GridRec and ML-EM. We use the distributed KMeans
implementation of MLlib and the GridRec, ML-EM of TomoPy. In the experiment we
utilize the MASS Mini-App with 1 node and 8 producer processes to continuously
produce messages of 0.3\,MB/5000 points for KMeans and 2 MB/1 point for the
light source scenarios. This way are able to simulate a complex read/write
workloads on the Kafka broker. We use 12 partitions/node for the Kafka topic.
The Mini-App uses Spark Streaming with a mini-batch window of 60\,sec.

\begin{figure}
	\centering
	\includegraphics[width=.49\textwidth]{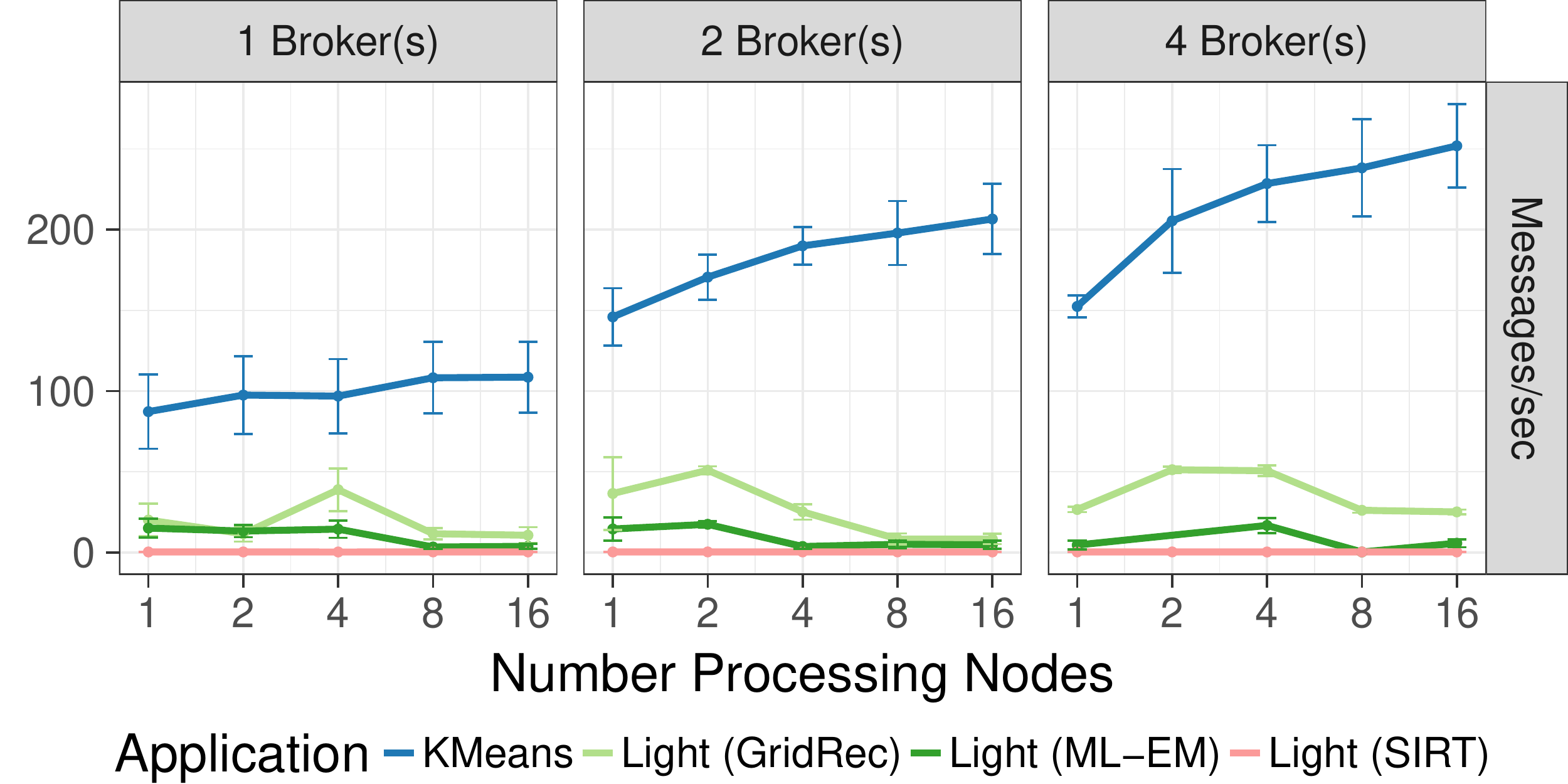}
	\caption{\textbf{MASA Throughput for KMeans and two Light Source Reconstruction 
		Algorithms:} KMeans scales well with increasing numbers of nodes. 
		GridRec shows a higher throughput than ML-EM as it less 
	computation complex. Scaling of both reconstruction algorithms is limited 
	by I/O contention.  \label{fig:MST_Processing_throughput} \upp\upp\upp\upp\upp\upp}
\end{figure}

Figure~\ref{fig:MST_Processing_throughput} shows the results of the experiment.
The processing throughput depends on various aspects, such as the bandwidth to
the message broker, computational complexity, and the scalability of the
processing algorithm. The KMeans application shows the highest throughput. It
scales both increasing number of processing nodes. For example, it is apparent
that in the 1 and 2 broker scenario, the I/O to the broker constraints the
performance. With additional broker nodes, the available bandwidth and
parallelism increasing. Spark Streaming assigns 1 task per Kafka partition.
This is visible in a significant increase in throughput. With KMeans we were
able to achieve a maximum throughput of 277 messages/sec and thus, were easily
able to sustain the generated data rate.

The throughput of the light source reconstruction algorithms is significantly
worse with maximum 63 message/sec for GridRec and 22 messages/sec for ML-EM. As
describe iterate algorithms, such as ML-EM are more demanding than GridRec.
Additional broker nodes yielded in significant performance improvements.
Additional processing nodes improved the performance as long the bandwidth to
the resource broker was able to keep up with the additional processing
resources. The amount of data transferred is with 2 MB/message significant
larger than in the KMeans scenario. Further, we observed some resource
contentions caused by running multiple instances of the algorithm on the same
node and the need to buffer a significant number of messages. The results show
the importance of resource management - only if the bandwidth and
read-parallelism to the data source or broker is large enough additional
compute resources are beneficial.

\emph{Discussion:} As demonstrated, the overhead for Pilot-Streaming is small: the startup time for dynamically starting Kafka, Dask and Spark clusters is
outweighed by the benefits of improved flexibility, resource isolation (per
application components), and the ability to scale components independently (at
runtime if needed). We demonstrated the scalability of the framework by
managing large streaming landscapes of Dask, Spark and Kafka concurrently on up 
to 32 nodes, 1536 virtual cores, and 4\,TB of memory  achieving throughputs of 
up to 390 MB/sec for the lightsource scenario. This throughput is large enough 
to sustain the LCLS-I data stream with a high enough sampling rate. At the 
current setup, the processing side is the bottleneck. We are only able to process a fraction of the data. Scaling stream processing is more difficult than
scaling batch analytics workload as it requires a careful balance of bandwidth
to/from the data source respectively the broker and compute resources. In
particular, it can be difficult to diagnose bottlenecks in the broker, as the
varying mixture of write/read I/O makes the performance often unpredictable.
Pilot-Streaming provides the necessary abstractions to manage resources
effectively at runtime on application-level.

The Streaming Mini-Apps simplify streaming application development
and performance optimizations. Using the Streaming Mini-Apps, we were
able to emulate various complex application characteristics. It is apparent
that the different frameworks and application components each have unique
scaling characteristics and resource needs. Even for optimization of just one
component a large number of combinations of experiments is required. On
streaming application-level this leads to a combinatorial explosion of
configurations. The Streaming Mini-Apps and Pilot-Streaming provide essential
tools for automating this process. In the future, we will use both frameworks
as foundation for higher-level performance optimization approaches, e.\,g.,
modeling the performance of each component, the usage of experimental design
and machine learning techniques for performance predictions.

\section{Conclusion and Future Work}
\upp

Pilot-Streaming fills an important gap in supporting stream processing on HPC
infrastructure by providing the ability to on-demand deploy and manage
streaming frameworks and applications. This capability is crucial for an
increasing number of scientific applications, e.\,g., light source sciences,
to generate timely insights and allow steering. The landscape of tools and frameworks for message
brokering, data storage, processing and analytics is diverse. Pilot-Streaming
currently integrates with Kafka, Spark Streaming, Dask and Flink. Its
flexible, plug-in architecture allows the simple addition of new frameworks.
Streaming applications can have unpredictable and often, external induced
resource needs, e.\,g. driven by the data production rate. Pilot-Streaming
addresses these needs with a well-defined resource model and abstraction that
allows the adjustments of the allocated resources for each component at runtime.
Another important contribution are the Streaming Mini-Apps, which
simplifies the development of streaming pipelines with the ability to emulate
data production and processing. We demonstrated the variety
of features of this framework with several experiments using a streaming KMeans
and different light source analysis algorithms.

This work represents the starting point for different areas of research: We
will extend Pilot-Streaming to support highly distributed scenarios enabling
applications to push compute closer to the edge for improved data
locality. The Streaming Mini-Apps will be the basis for the
development and characterization of new streaming algorithms, e.\,g.\
additional reconstruction algorithms and deep learning based object
classification algorithms. We will explore the usage of accelerators
(such as GPUs) to support compute-intensive deep learning workloads. Another area of research are steering capabilities. Further, we will
continue to utilize the Streaming Mini-Apps to improve our understanding of
streaming systems and embed this into performance models that can inform
resource and application schedulers about expected resource needs.

\vspace{2mm}

\scriptsize
{\bf Acknowledgements:} We thank Stuart Campbell and Julien Lhermitte
(BNL) for guidance on the light source application. This work is funded by NSF 1443054 and 1440677. Computational resources were provided by NSF XRAC award TG-MCB090174.

%
%
%
%
%
%
%
%
%
%
%
%
%
%
%
%
%
%
%
%
%
%
%

%
%
%
%
%
%
%
%
%

%
%
%
%

\end{document}